\documentclass[conference]{IEEEtran}
\IEEEoverridecommandlockouts

\usepackage{cite}
\usepackage{amsmath,amssymb,amsfonts}
\usepackage{subfigure}
\usepackage{svg}
\usepackage{booktabs}
\usepackage{graphicx}
\usepackage{multicol}
\usepackage{multirow}
\usepackage{hyperref}
\usepackage{tabularx}
\usepackage{booktabs}
\usepackage[noindentafter]{titlesec}
\usepackage{dsfont}
\usepackage{algorithm}
\usepackage[noend]{algpseudocode}
\usepackage{mathtools, nccmath}
\usepackage{dblfloatfix}
\usepackage{amsmath}
\usepackage{wrapfig}
\usepackage{float}
\usepackage{mathtools}
\usepackage{enumitem}
\usepackage{flushend}
\usepackage{xcolor}
\newcommand{\blue}[1]{\textcolor{black}{#1}}

\graphicspath{ {images/} }
\usepackage{amsmath}
\usepackage{amssymb}

\def\BibTeX{{\rm B\kern-.05em{\sc i\kern-.025em b}\kern-.08em
    T\kern-.1667em\lower.7ex\hbox{E}\kern-.125emX}}
    
\begin{document}

\title{CAROL: Confidence-Aware Resilience Model for Edge Federations}
\author{
 \IEEEauthorblockN{Shreshth Tuli\IEEEauthorrefmark{1}, Giuliano Casale\IEEEauthorrefmark{1} and Nicholas R. Jennings\IEEEauthorrefmark{1}\IEEEauthorrefmark{2}}
\IEEEauthorblockA{{\IEEEauthorrefmark{1}Department of Computing, Imperial College London, UK}}
 \IEEEauthorblockA{{\IEEEauthorrefmark{2}Loughborough University}}
 \IEEEauthorblockA{Emails: \{s.tuli20, g.casale\}@imperial.ac.uk, n.r.jennings@lboro.ac.uk}\protect %
}

\urlstyle{tt}
\maketitle
\thispagestyle{plain}
\pagestyle{plain}


\begin{abstract}
In recent years, the deployment of large-scale Internet of Things (IoT) applications has given rise to edge federations that seamlessly interconnect and leverage resources from multiple edge service providers. The requirement of supporting both latency-sensitive and compute-intensive IoT tasks necessitates service resilience, especially for the broker nodes in typical broker-worker deployment designs. Existing fault-tolerance or resilience schemes often lack robustness and generalization capability in non-stationary workload settings. This is typically due to the expensive periodic fine-tuning of models required to adapt them in dynamic scenarios. To address this, we present a confidence aware resilience model, CAROL, that utilizes a memory-efficient generative neural network to predict the Quality of Service (QoS) for a future state and a confidence score for each prediction. Thus, whenever a broker fails, we quickly recover the system by executing a local-search over the broker-worker topology space and optimize future QoS. The confidence score enables us to keep track of the prediction performance and run parsimonious neural network fine-tuning to avoid excessive overheads, further improving the QoS of the system. Experiments on a Raspberry-Pi based edge testbed with IoT benchmark applications show that CAROL outperforms state-of-the-art resilience schemes by reducing the energy consumption, deadline violation rates and resilience overheads by up to 16, 17 and 36 percent, respectively.
\end{abstract}


\maketitle




\section{Introduction}

The fifth industrial revolution, named Industry 5.0, marks a significant shift in the technological backbone of industrial applications such as the Internet of Things (IoT) and Artificial Intelligence (AI)~\cite{maddikunta2021industry}. It allows end-to-end integration of sensors and actuators close to the user with geographically distributed servers via a large number of intermediary smart edge nodes~\cite{zanella2014internet}. Such frameworks enable the deployment of latency-critical and compute-intensive AI-based IoT applications on edge devices for low response time and large-scale service delivery. It achieves this by the collective adoption of edge federations that bring together the software, infrastructure and platform services of multiple edge computing environments. Unifying the computational devices of multiple service providers enables such paradigms to accommodate sudden spikes in user demands~\cite{rochwerger2009reservoir}. However, due to the distributed nature of such environments, centralized management of their resources is susceptible to service downtimes in high workload settings. Thus, most edge federations employ a broker-worker topology with multiple broker nodes managing the system~\cite{tuli2019fogbus}. Here, the brokers receive tasks from the users and delegate processing to one of the worker nodes within their control, referred to as local edge infrastructures (LEIs). This leads the processing bottlenecks in such systems, \textit{i.e.}, the broker nodes, play a vital part in resilient service delivery.  

\textbf{Challenges.} \blue{The key challenge addressed in this paper is dealing with the diverse effects of byzantine node failures, such as increased task response time and violations of Service Level Objectives (SLOs), requiring different remediation steps to maintain system performance and reduce service downtime. This is motivated by the excessive load that AI-based IoT applications put on the limited computational capacity and working memory (up to 8GB) of edge devices~\cite{hao2018edge}.} Higher computational load translates to increased task processing times and SLO violation rates. Excessive memory load is typically dealt with using storage mapped virtual memory, which in most settings is a network-attached disk~\cite{shao2020communication}. This also leads to higher response times due to time-consuming data transfers over congested backhaul edge networks~\cite{bilal2019collaborative}. All these issues lead to persistent resource contention and faulty behavior that adversely affects system QoS. Furthermore, without resource intensive cloud backends to rely on in edge federations, it becomes challenging to deploy modern deep learning based management solutions in resource constrained edge devices.
%
In typical broker-worker federations, if a worker node fails, a broker could act as a worker and complete the task or allocate the same task to another worker~\cite{tuli2019fogbus, tuli2021gosh, sajjad2016spanedge}. However, if a broker fails, all active tasks within the LEI and all incoming tasks that are sent to the broker are impacted. This makes broker resilience crucial in large-scale deployments. \blue{However, the problem of broker resilience is hard as maintaining hot-redundancy by replicating running task instances makes edge nodes more susceptible to failures~\cite{li2020hotdedup}. The difficulty primarily arises from the critical need for result delivery with low latency and high accuracy~\cite{ahuja2012survey}. A vital parameter that controls this is the number of brokers in the system. For a fixed number of devices in the system, a high broker count translates into having fewer workers, reducing the average throughput of the system. Low broker count can cause bottlenecks and contentions, increasing fault frequency.} So we need to consider both the increase or decrease of the number of brokers in the system.  Furthermore, the statistical moments and correlations of the workload characteristics are non-stationary and vary over time, requiring continuous steps to adapt management models. This imposes additional overheads and impacts QoS~\cite{ristov2020resilient}.

\begin{table*}[]
    \centering
    \caption{Comparison of related works with different parameters (\checkmark means that the corresponding feature is present).}
    \label{tab:related_works}
    \begin{tabular}{@{}lccccccccc@{}}
    \toprule
    \multirow{2}{*}{Work} & \multirow{2}{*}{IoT} & \multirow{2}{*}{Approach} & Consider Broker & QoS & \multicolumn{5}{c}{Performance parameters}\tabularnewline
    \cmidrule{6-10}
     &  &  & Resilience & Prediction & Energy & Response Time & SLO Violations & Overheads & Memory\tabularnewline
     \midrule
    DYVERSE \cite{dyverse} & \checkmark & Heuristic & \checkmark &  & \checkmark &  & \checkmark & \checkmark & \tabularnewline
    DISP \cite{rathod2019scalability} &  & Heuristic &  &  &  & \checkmark & \checkmark &  & \tabularnewline
    LBM \cite{khattak2019utilization} & \checkmark & Heuristic & \checkmark & \checkmark & \checkmark &  &  &  & \tabularnewline
    FDMR \cite{fdmr} &  & Meta-Heuristic &  &  &  & \checkmark & \checkmark &  & \tabularnewline
    ECLB\cite{eclb} & \checkmark & Meta-Heuristic & \checkmark & \checkmark & \checkmark &  & \checkmark &  & \tabularnewline
    LBOS \cite{lbos} & \checkmark & RL &  & \checkmark & \checkmark & \checkmark &  & \checkmark & \checkmark\tabularnewline
    ELBS \cite{elbs} & \checkmark & Surrogate Model & \checkmark & \checkmark & \checkmark & \checkmark & \checkmark &  & \tabularnewline
    FRAS~\cite{fras} &  & Surrogate Model &  & \checkmark & \checkmark &  & \checkmark &  & \checkmark\tabularnewline
    TopoMAD~\cite{topomad} &  & Reconstruction &  & \checkmark & \checkmark & \checkmark & \checkmark &  & \tabularnewline
    StepGAN~\cite{stepgan} & \checkmark & Reconstruction &  & \checkmark & \checkmark & \checkmark & \checkmark &  & \tabularnewline
    \textbf{CAROL} & \checkmark & Surrogate Model & \checkmark & \checkmark & \checkmark & \checkmark & \checkmark & \checkmark & \checkmark\tabularnewline
    \bottomrule
    \end{tabular}
\end{table*}

\textbf{Existing solutions.} Many recent works~\cite{dyverse, lbos, elbs, fras} aim to provide effective fault tolerance or system resilience by leveraging heuristics, meta-heuristics or AI models. The heuristic and meta-heuristic methods often perform poorly in dynamic settings where workload characteristics and SLO demands are non-stationary~\cite{elbs,fras}. AI-based methods typically utilize reinforcement learning (RL) or neural networks as surrogate models to predict future system states and estimate their QoS. \blue{The predicted QoS values indicate the chances of future broker or worker breakdowns and proactively manage the broker-worker topology to avoid service downtimes.} These methods leverage fault-aware scheduling~\cite{eclb}, preemptive migration based load-balancing~\cite{lbos} or auto-scaling techniques~\cite{fras}. RL or surrogate models trained on large datasets enable such techniques to be effective even in unseen settings by exploration and neural network fine-tuning. This is a process where the model utilizes the data generated during execution to update the neural network parameters in an online fashion. However, these solutions have several limitations. Most AI-based approaches are designed for cloud setups with GPUs for faster training of the underlying neural networks and result generation~\cite{roopaei2017deep}. To deploy these models in resource-constrained edge settings, various model compression and parameter neural network splitting are required, adversely impacting their performance~\cite{shao2020communication}. This also impacts the periodic model fine-tuning, increasing the training times in distributed neural network settings~\cite{luping2019cmfl}. Together, these factors mean that neural network fine-tuning process consumes large portions of the computational and memory resources of edge devices. This impacts the execution of the management tasks running in the broker nodes, further leading to contentions in broker nodes. To resolve this, there is a need for a lightweight solution that parsimoniously fine-tunes AI methods. 

\textbf{Key insights and our contributions.} For a lightweight broker resilience model, we develop a method that uses a neural network to predict system QoS and also indicates when to fine-tune the network to adapt to non-stationary settings. The novel insight is using a generative network as a surrogate model to optimize QoS. Unlike prior work that use traditional feed-forward or recurrent neural networks to predict QoS~\cite{eclb, elbs}, using specific generative models not only allows us to estimate QoS of future states, but also an indicator of the prediction confidence. Examples of such a generative network with low memory footprint are recently-proposed models by the AI theory community such as Generative Optimization Networks (GONs)~\cite{tuli2021generative} or SpareVAEs~\cite{ashman2020sparse}, which provide an alternative to Generative Adversarial Networks (GANs)~\cite{goodfellow2014generative} or Variational Auto-encoders (VAEs)~\cite{kingma2013auto} with much lower memory footprint and therefore suitable for edge execution. Models with a reduced footprint of this kind pave the way to the use of GAN-type methods in edge devices with resource footprint constraints. \blue{In our recent work~\cite{dragon}, we have demonstrated an application of GONs to decentralized fault-tolerance finding up to 82\% improvement in service-level compliance when a local edge infrastructure runs a GON in its broker to detect faults within its edge devices. Contrary to that work, we focus here instead on the application of GONs to the problem of broker resilience, which is not touched upon in \cite{dragon}.} 

In the proposed work, an \blue{offline} trained GON model with \blue{labelled} data generated from a system with similar behavior as that of the training dataset would typically give high confidence scores, whereas as the system behavior changes, the confidence score declines (more details in Section~\ref{sec:method}). Dynamic thresholding techniques allow us to decide confidence thresholds below which we fine-tune the network parameters. Thus, we only run fine-tuning measures when required, significantly reducing overheads and improving system QoS. In this work we present CAROL: \underline{C}onfidence \underline{A}ware \underline{R}esilience M\underline{o}de\underline{l} for edge federations. CAROL is the first system that uses a GON in edge brokers to reactively run topology optimization to optimize system QoS. The GON model is trained using log traces on DeFog benchmarks~\cite{mcchesney2019defog}. Extensive experiments on a Raspberry-Pi based federated edge cluster show that CAROL performs {best} in terms of QoS metrics. To test the generalization of the model, we test on different benchmarks as workloads, namely AIoT~\cite{luo2018aiot}. Specifically, CAROL reduces energy consumption, SLO violation rates and resilience overheads by up to 16, 17 and 36 percent, respectively, compared to state-of-the-art baselines. 


\section{Related Work}
\label{sec:related_work}

We list in Table~\ref{tab:related_works} the prior work, dividing such methods into two classes: heuristic and meta-heuristic methods (rows 1-5 of Table~\ref{tab:related_works}) and AI-based methods (rows 6-10). Many of these methods use simple strategies to deal with broker failures or are unable to efficiently adapt in non-stationary environments.

\textbf{Heuristic and Meta-heuristic methods.} Most fault prevention techniques for cloud and edge computing employ some form of heuristics or meta-heuristic approaches. Methods like DYVERSE~\cite{dyverse} use dynamic vertical scaling in multi-tenant edge systems to manage resources assigned to IoT applications to improve scalability. It uses an ensemble of three heuristics (system-aware, community-aware and workload-aware) to dynamically allocate priority scores to the active applications in the system. DYVERSE uses AI-based face detection and online game workload to validate in a controlled edge computing environment. However, for broker failures, it allocates the worker with the least CPU utilization as the next broker of the same LEI. Federated Distributed MapReduce (FDMR)~\cite{fdmr} is another MapReduce based framework that uses integer linear programming for fault-tolerant distributed task scheduling. Another approach, namely Distributed IoT Service Provisioning (DISP)~\cite{rathod2019scalability} technique, balances load across edge and fog nodes by comparing CPU utilization and response time metrics for each node. However, due to the modeling limitations, such methods do not scale for real-time operations, making them unsuitable for mission-critical edge applications. A similar approach, Load Balancing Mechanism (LBM)~\cite{khattak2019utilization}, executes user requests within edge nodes and utilizes metrics like network traffic and CPU utilization to decide the worker node that takes over as a broker in case of a failure.  However, this work assumes a homogeneous edge setup and has been shown to often perform poorly in heterogeneous environments~\cite{nezami2021decentralized}.  The Energy-efficient Checkpointing and Load Balancing (ECLB)~\cite{eclb} technique uses Bayesian methods to classify host machines into three categories: overloaded, underloaded and normal execution. This classification is used to decide appropriate task migrations to reduce the number of overloaded hosts. However, this model only considers computational overloads and does not consider other fault types like node thrashing or network failures that could lead to broker nodes being compromised. We use the best performing methods, DYVERSE and ECLB, as baselines in our experiments, as demonstrated in prior work~\cite{dyverse, eclb}.

\textbf{AI-based methods.} Recently, several resilience models have been proposed that leverage AI methods like RL, surrogate or reconstruction modeling. An RL based approach is Load Balancing and Optimization Strategy (LBOS)~\cite{lbos} that allocates the resources using RL. The reward of the RL agent is calculated as a weighted average of multiple QoS metrics to avoid system contention by balancing the load across multiple compute nodes. The values of the weights are determined using genetic algorithms. LBOS observes the network traffic constantly, gathers the statistics about the load of each edge server, manages the arriving user requests and uses dynamic resource allocation to assign them to available edge nodes. However, RL approaches are known to be slow to adapt in dynamic settings~\cite{tuli2021cosco}. Most other approaches use neural networks as a surrogate model. For instance, Effective Load Balancing Strategy (ELBS)~\cite{elbs} is a recent framework that offers an execution environment for IoT applications and creates an interconnect among cloud and edge servers. The ELBS method uses the priority scores to proactively allocate tasks to edge nodes or worker nodes as brokers to avoid system failures. It uses a fuzzy inference system to calculate the priority scores of different tasks based on three fuzzy inputs: SLO deadline, user-defined priority, and estimated task processing time. The priority values are generated by a neural network acting in the capacity of a surrogate of QoS scores.  Another similar method is the Fuzzy-based Real-Time Auto-scaling (FRAS)~\cite{fras} technique that leverages a virtualized environment for the recovery of IoT applications that run on compromised or faulty edge nodes.  Here, FRAS executes each IoT application in a virtual machine (VM) and performs VM autoscaling to improve execution speed and reduce execution costs. The VM autoscaling decisions making involves inference of system QoS using a fuzzy recurrent neural network as a surrogate model. A major drawback of such surrogate modeling methods is that their parameters need to be periodically fine-tuned to adapt to dynamic environments, giving rise to high overheads. Other methods in this category generate a reconstruction of the system state and use the deviation from the input to indicate the likelihood that the state is faulty. For instance, TopoMAD~\cite{topomad} uses a topology-aware neural network that is composed of a Long-Short-Term-Memory (LSTM) and a variational autoencoder (VAE) to detect faults. However, the reconstruction error is only obtained for the latest state, limiting them to using reactive fault recovery policies. Other methods use slight variations of LSTM networks with either dropout layers~\cite{girish2021anomaly, tuli2020modelling}, causal Bayesian networks~\cite{gan2020sage} or recurrent autoencoders~\cite{chouliaras2021detecting}. A GAN-based approach that uses a stepwise training process, StepGAN~\cite{stepgan}, converts the input time-series into matrices and executes convolution operations to capture temporal trends. These methods use various thresholding techniques like Peak Over Threshold (POT)~\cite{siffer2017anomaly} or Kernel Density Estimation (KDE)~\cite{martin1996non}. However, such techniques are not agnostic to the number of hosts or workloads as they assume a maximum limit of the active tasks in the system. Moreover, even though more accurate than heuristic based approaches, deep learning models such as deep autoencoders, GANs and recurrent networks are adaptive and accurate, but have a high memory footprint that adversely affects system performance. From this category, we test the above mentioned approaches on the testbed described in Section~\ref{sec:experiments} and use the empirically best techniques based on our experiments as baselines: LBOS, ELBS, FRAS, TopoMAD and StepGAN.

\section{Methodology}
\label{sec:method}

\begin{figure*}
    \centering
    \includegraphics[width=\textwidth]{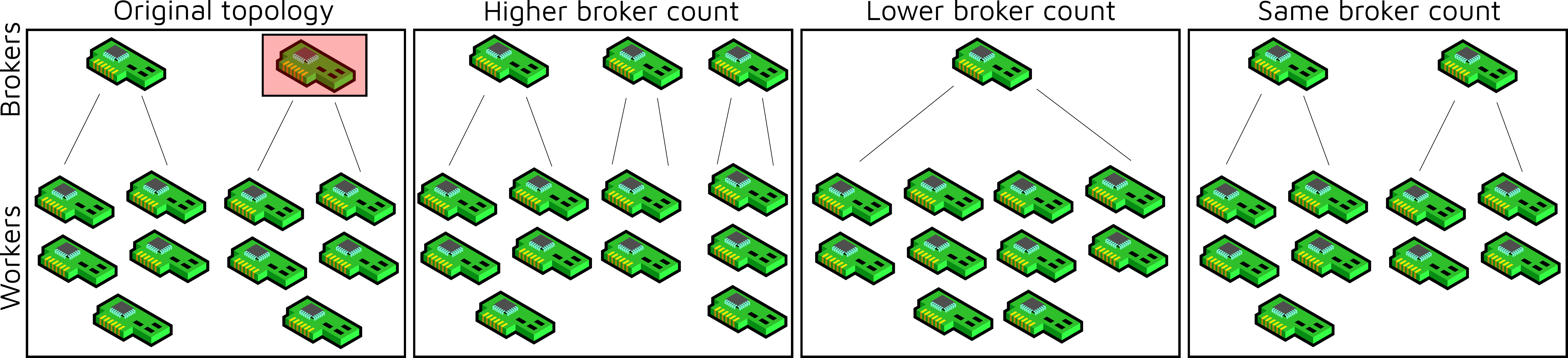}
    \caption{Possible node-shifts in CAROL. Nodes marked in red are the high consumption brokers that break-down.}
    \label{fig:scale}
\end{figure*}

\subsection{Environment Assumptions and Problem Formulation}
\label{sec:systemmodel}

\textbf{System Model.} As is commonplace in federated edge computing environments, we assume a system with heterogeneous nodes configured in a broker-worker fashion~\cite{weissman1996federated}. The assignment of edge nodes as brokers or workers and the allocation of all workers to one of a broker defines the topology of the system. We denote the number of edge nodes by $H$. As is common in edge federations, we assume that the broker nodes of LEIs are interconnected and we allow data sharing among brokers to facilitate transferring management tasks of workers across LEIs. We also assume a bounded timeline of execution of tasks within the federated environment, which we divide into fixed-sized scheduling intervals, where $I_t$ denotes the $t$-th interval ($t$ ranging from 0 to $T$). We also consider that the edge broker can sample the resource utilization metrics of all hosts within its LEI group at any time including CPU, RAM, disk/network bandwidth, some additional fault-related metrics including consumption of the swap space, disk buffers, network buffers, disk and network I/O waits.

\textbf{Fault Model.} As per prior work~\cite{ye2018fault, dftm}, we consider a byzantine failure model for the edge nodes in our setup~\cite{driscoll2003byzantine}. We assume that all failures are recoverable, \textit{i.e.}, the machines can be rebooted to resume execution from their last working state\blue{, by frequently updating snapshots of the active hosts in the system.} As in prior work, edge hosts in the same LEI are connected to the same power supply and unrecoverable faults like outages are ignored~\cite{dftm}. Failures can occur for a variety of reasons, including software or hardware defects, such as an infinite loop or long-running uninterruptible computation, resource exhaustion (thrashing), under-performing hardware (throttling), external events such as a slow computer network, misconfiguration, and compatibility problems. We realize this using an existing fault injection module~\cite{ye2018fault} to create different fault types like CPU overload, RAM contention, Disk attack and DDOS attack. We specifically restrict our attention to faults that manifest in the form of resource over-utilization; for instance, a DDOS attack could lead to contention at the network interface. A broker or worker node may become unresponsive due to this resource over-utilization~\cite{vasilakos2020towards}. This work aims to find the best system topology in terms of estimated QoS in case one or more broker nodes fail. This is crucial as broker failures lead to service downtimes from all nodes in the LEI. In case of worker failures, we simply rerun tasks on the worker with the least resource utilization in the LEI. As this work takes a step in the complementary direction of traditional load-balancing and autoscaling methods, existing fault-tolerance techniques may be leveraged for worker management~\cite{eclb, dyverse}.

\textbf{Workload Model.} All the tasks in our system are generated by users and transferred to the edge federation using gateway devices. We assume that gateway devices send tasks to the closest broker in terms of network latency, breaking ties uniformly at random. We assume a bag-of-tasks workload model, where a set of independent tasks enter each LEI at the start of each scheduling interval. Each task has an associated (soft) SLO deadline. 

\textbf{Formulation.} In this work, we assume that the edge brokers run software solutions to manage their worker nodes. This includes making scheduling decisions for all incoming and active tasks in the system. Such solutions may include various fault tolerance and resilience models. We denote the set of brokers by $B$ and workers by $W$. We also refer to these as broker and worker layers as per prior work~\cite{tuli2019fogbus}. In this work, in each scheduling interval, we check which broker or worker nodes are unresponsive (inactive) and update the system topology to optimize QoS. We encode the undirected topology graph of the federated edge environment at the start of the interval $I_t$ as $G_{t-1}$. Formally, at the start of the interval $I_t$, all active nodes in the system are utilized to generate the topology $G_t$ to execute tasks in $I_t$. The new graph $G_t$ may be the same as $G_{t-1}$ in case the active node set is unchanged.

To generate the graph $G_t$, we utilize the performance metrics of the previous interval, denoted by $M_{t-1}$, and the scheduling decision by $S_t$. The performance metrics include resource utilization and QoS metrics such as energy consumption and response time. For $S_t$ we assume an underlying scheduler in the system independent from the proposed fault-tolerance solution. Using the input graph topology $G_t$, performance metrics $M_{t-1}$ and an input scheduling decision $S_t$, the model needs to predict the performance metrics for the current interval, \textit{i.e.}, $M_t$ and a confidence score $C_t$. Using this model, we optimize over the topology space to find the optimal $G_t$ for interval $I_t$. Without loss in generality, whenever unambiguous, we drop the subscripts for the sake of simplicity. Hence, we only refer to the inputs and outputs as $G$, $M$, $S$ and $C$.

\subsection{CAROL Model}
\label{sec:model}
\textbf{Node-Shift.} We assume that the network has tens of broker and worker nodes. In our work, the number of brokers and the system topology is known to all nodes in the federation. Whenever a broker node fails, we consider the worker nodes of the corresponding broker as being ``orphaned''. In such an event, either one of the worker nodes is shifted to the broker layer, or another node in the broker layer manages the orphaned nodes. This shift of nodes from worker layer to broker layer gives rise to the name ``node-shift'' and is similar to the fault-tolerance schemes used in software defined networks (SDNs)~\cite{samarji2021fault}. Considering an input topology $G$, there are multiple types of node shifts that may increase, decrease or even keep the broker count static. Three types of \textit{worker-to-broker} node-shift mechanisms considered in this work are described below with a visualization presented in Figure~\ref{fig:scale} instead. We similarly consider the respective counterparts as the \textit{broker-to-worker} node-shifts.
\begin{itemize}
    \item \textit{Type 1:} If a broker node fails, two of the orphaned nodes may be shifted to the broker layer and the remaining orphaned nodes may be evenly distributed among these two new brokers. This increases the broker count by one, also increasing the management capacity of the system for higher throughput at the broker layer.
    \item \textit{Type 2:} The orphaned nodes could be assigned to another active broker in the system. This decreases the broker count than before failure and increases the computational throughput of the worker layer.
    \item \textit{Type 3:} One of the orphaned worker nodes may be assigned to be the broker for the other worker nodes. This keeps the same number of broker nodes as before the failure. The three node-shift types may be utilized to trade off the throughputs of the broker and worker layers. 
\end{itemize}

However, in heterogeneous edge federations, where nodes with disparate resource capacities may be present, the choice of the worker node to shift to the broker layer becomes vital. Moreover, with heterogeneous broker nodes, the choice of the broker nodes to which the orphaned nodes are assigned affects the QoS of the system. Another crucial factor is the overhead corresponding to the node-shift operations. This is due to the initialization of the broker management systems and synchronization of the system topology with other brokers in the federation.  

Another critical aspect in finding the best graph topology is the workload heterogeneity across broker nodes. As shown in Figure~\ref{fig:scale}, the node operations can lead to a different number of worker nodes and subsequently computational throughput of disparate LEIs. A single node-shift step might not be sufficient to manage resources in case of different computational loads across LEIs. Thus, a sequence of node-shifts may need to be tested to discover the topology with optimal QoS. However, generating QoS scores for a large number of node-shift sequences may not be feasible in latency-critical settings where generating QoS scores by execution or simulation might be a time-consuming activity. \blue{This is because a node-shift entails transfer of broker level data to the worker node and initializing management software containers, both of which give rise to a higher latency.}

\textbf{Confidence-Aware Model.} To eschew the costly execution of multiple node-shift sequences and observe their effects on QoS, we need a lightweight model that mimics the behavior of a computationally expensive simulation or execution~\cite{kochenderfer2019algorithms}. Such models are referred to as ``surrogate models'' in the literature~\cite{kochenderfer2019algorithms}.  Using a surrogate model enables such methods to run optimization in the input space and generate decisions, such as node-shift sequences that give $G_t$, giving high estimated QoS. Recent surrogate optimization techniques use gradient-based optimization compared to evolutionary search strategies that facilitate quick convergence~\cite{tuli2021cosco}. However, in discrete search spaces, such as graph topologies, such techniques typically select the discrete point close to the converged point in the continuous space. This may often lead to non-optimal solutions~\cite[\S 19.1]{kochenderfer2019algorithms}. Another drawback of such methods is that the parameters of the neural networks need to be periodically fine-tuned, leading to high overheads~\cite{tuli2020dynamic}. To reduce such overheads, it is crucial to fine-tune models only when the system or workload configurations change. To do this, we take motivation from confidence-aware deep learning~\cite{moon2020confidence} and a recently proposed class of generative models, called Generative Optimization Networks (GONs)~\cite{tuli2021generative}, to integrate in CAROL a memory efficient QoS surrogate. This is used to predict a confidence score, such that for low confidence the model can be fine-tuned with online generated data.

\textbf{GON based Neural Network.} GANs are based on a pair of neural networks, a generator and a discriminator where the generator takes random noise samples and outputs samples from a data distribution as inputs. The discriminator predicts a likelihood score for the input belonging to the target distribution. Unlike traditional GANs, we leverage a memory-efficient Generative Optimization Network (GON)~\cite{tuli2021generative}. GONs are similar to GANs, but do not use the generator, significantly reducing the memory footprint of the neural network. We now describe the working of the GON model and confidence-based QoS prediction.

Consider a discriminator network $D$ with parameters $\theta$ that takes as input graph topology $G$, performance metrics $M$ and scheduling decision $S$. The output of $D$, \textit{i.e.}, $D(M, S, G; \theta)$ is a likelihood score of $G$, $M$ and $S$ belonging to a distribution. When trained for a dataset generated from the normal execution of an edge federation, the GON model predicts a high score for an input tuple $(M, S, G)$ in the distribution of the dataset and a low score for an unseen tuple. This allows us to translate the likelihood score output as a measure of the confidence of the network. Essentially, for dynamic systems, when the confidence score drops below a threshold, it indicates us to fine-tune the model. This enables parsimonious model fine-tuning, giving us an effective compromise between prediction performance and training overheads. 

To generate the performance scores for an input graph topology $G$, we can randomly initialize $M$ and maximize the discriminator output by ascending the stochastic gradient
\begin{equation}
\label{eq:opt}
    M \gets M + \gamma \nabla_M \log \big( D(M, S, G; \theta) \big),
\end{equation}
where $\gamma$ denotes the step size in the optimization loop. We use log-likelihood instead of likelihood scores for training stability~\cite{goodfellow2014generative}. Thus, starting from an input $(S, G)$ pair, the above optimization loop converges us to give performance metrics $M^*$ and a confidence score of $D(M^*, S, G)$. 

\begin{algorithm}[!t]
    \begin{algorithmic}[1]
    \Require Dataset $\Lambda = \{M_t, S_t, G_t\}_{i=0}^T$
    \For{ number of training iterations }
    \State Sample minibatch of size $m$ performance metrics $\{M^{(1)}, \ldots, M^{(m)}\}$ from $\Lambda$.\label{line:sample_data}
    \State Sample minibatch of size $m$ noise samples $\{Z^{(1)}, \ldots, Z^{(m)}\}$.\label{line:sample_noise}
    \State Generate new samples $\{Z^{*(1)}, \ldots, Z^{*(m)}\}$ by running the following till convergence
    \begin{equation*}
        Z \gets Z + \gamma  \cdot \nabla_{Z}\log \big( D(Z, S, G; \theta) \big).%
    \end{equation*}\label{line:optimize}\vspace{-10pt}
    \State Update the discriminator by ascending the stochastic gradient.
    \begin{equation*}
    \resizebox{0.94\hsize}{!}{%
        $\displaystyle \nabla_\theta \frac{1}{m}\! \sum_{i=1}^m\! \big[\log\! \big(D(M^{(i)}, S, G; \theta) \big) + \log\! \big(1 - D(Z^{*(i)}, S, G; \theta) \big) \big].$}
    \end{equation*} \label{line:train}
    \EndFor
    \end{algorithmic}
\caption{Minibatch stochastic gradient based training of GON model in CAROL. Input is dataset $\Lambda$ and hyperparameters $m$ and $\gamma$. }
\label{alg:gon}
\end{algorithm}

\textbf{Offline Model Training.} To train the GON model $D$, we first collect an execution trace $\Lambda = \{M_t, S_t, G_t\}_{i=0}^T$. A summary of the training process is presented in Algorithm~\ref{alg:gon}. We leverage an adversarial training process where we use the generated data (say $Z^*$) as fake samples and datapoints from $\Lambda$ (say $M$) as real samples and train the model using the binary cross-entropy loss
\begin{equation}
    L = \log \big(D(M, S, G; \theta) \big) + \log \big(1 - D(Z^*, S, G) \big),
\end{equation}
where $(M, S, G) \in \Lambda$. Thus, the model is trained to optimize the likelihood scores for fake and real samples in an adversarial fashion using the same loss function. This adversarial style training has a two-fold goal. First, the model learns to generate new samples ($Z^*$) such that the tuple $(Z^*, S, G)$ belongs to the data distribution. Thus, with sufficient training, converged $Z^*$ are close estimates of the performance metrics for inputs $G$ and $S$. Second, for previously seen datapoints, \textit{i.e.}, $(M, S, G)$ tuples, the predicted confidence scores are high. If the $(S, G)$ input is unseen during training, then the converged confidence score $D(Z^*, S, G)$ would also be low. In practice, for the scheduling interval $I_t$, instead of starting from a random noise sample $Z$, we initialize $M$ as $M_{t-1}$ and then converge $D(M, S_t, G_t)$ to $M_t$. This exploits the temporal correlation between subsequent system states, facilitating quick convergence of~\eqref{eq:opt}.

\begin{algorithm}[t]
    \begin{algorithmic}[1]
    \Require
    \Statex Trained GON Model $D$; Running dataset $\Gamma$
    \Statex Broker set $B$; Worker set $W$
    \State \textbf{for} $t = 0 \text{ to } T$ \textbf{do}
    \State \hspace{0.8em} $S_t$ from underlying scheduler \label{line:schedulingdecision}
    \State \hspace{0.8em} $M_t$ from edge system \label{line:metrics}
    \State \hspace{0.8em} $G_t \gets \mathrm{NodeShift}(G_{t-1})$ \Comment{Initialize topology} \label{line:init}
    \State \hspace{0.8em} \textbf{for each} broker $b \in B$ \textbf{do}
    \State \hspace{1.5em} \textbf{if} $b$ fails \textbf{do}
    \State \hspace{2.2em} $G \in N(G_t, b)$ \label{line:init2}  \Comment{Random node-shift}
    \State \hspace{2.2em} $G_t \gets \mathrm{TabuSearch}(G, \Omega)$ \Comment{Tabu Search} \label{line:tabu}
    \State \hspace{0.8em} \textbf{if} no broker in $B$ fai \textbf{do}
    \State \hspace{1.5em} Add datapoint $(M_t, S_t, G_t)$ to $\Gamma$ \Comment{Save datapoint} \label{line:save}
    \State \hspace{0.8em} $C \gets D(M_t, S_t, G_t; \theta)$ \Comment{Confidence score} \label{line:confidence}
    \State \hspace{0.8em} $POT \gets \mathrm{PeakOverThreshold}(C)$ \Comment{POT calculation} \label{line:pot}
    \State \hspace{0.8em} \textbf{if} $C < POT$ \textbf{do} \label{line:migrate2}
    \State \hspace{1.5em} $L = \log (D(M, S, G; \theta) ) + \log (1 - D(Z*, S, G) )$\label{line:tune1}%
    \State \hspace{1.5em} Fine-tune $D$ using $L$ and \texttt{Adam} \label{line:tune2} \Comment{Fine-tune}
    \State \hspace{1.5em} Clear $\Gamma$ \label{line:clear} 
    \State \hspace{0.8em} Configure topology as $G_t$
    \State \hspace{0.8em} Schedule tasks using $S_t$
    \end{algorithmic} 
\caption{The CAROL resilience model.}
\label{alg:carol}
\end{algorithm}

\textbf{Finding optimal edge topology.}
\label{sec:fog-topo}
Having a surrogate model, at each scheduling interval $I_t$, we can now run optimization of the QoS by taking a convex combination of the QoS metrics in $M_t$. We denote this objective function as $O(M_t)$. Now, if a node  fails in $I_{t-1}$, then we can create the graph topology $G_t$ by a node-shift operation on $G_{t-1}$ (line~\ref{line:init} in Alg.~\ref{alg:carol}). However, to decide the optimal node-shift operation, we can run a local search in the topology space. We select the tabu search algorithm due to its deterministic nature and empirically faster convergence for the specific optimization problem we consider~\cite{connor2000comparison}. Thus, in case of a node failure, we start from graph $G_{t-1}$, apply a random node shift to generate $G$ and use tabu search to generate the graph $G_t$ such that the QoS score is optimized (line~\ref{line:tabu} in Alg.~\ref{alg:carol}). We perform this iteratively for each failed broker node to achieve the final graph topology. If no broker failed in $I_{t-1}$, then we add the datapoint to a running dataset $\Gamma$ (line~\ref{line:save} in Alg.~\ref{alg:carol}).

The local neighbors of a graph $G$ for a failed node, say $b$, is obtained by performing all possible node-shift operations. We denote this neighbourhood set of $G$ by ${N}(G, b)$ (line~\ref{line:init2} in Alg.~\ref{alg:carol}). The QoS score is obtained using the GON model described previously, where we evaluate $M_t$ using~\eqref{eq:opt} over $D(M, S_t, G)$ and use the objective function $O(M_t)$ for our local search. For notational convenience, we define the objective score of a graph $G$, parameterized by the GON model $D$, scheduling decision $S_t$ and objective function $O$ as $\Omega(G; D, S_t, O)$. As the network size increases, the number of possible node-shift operations also increases. Thus, we use a tabu list of a fixed size $L$ in our search scheme.

\begin{figure}
    \centering
    \includegraphics[width=\linewidth]{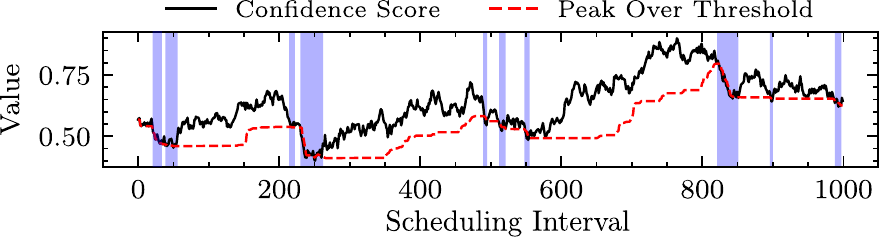}
    \caption{Visualizing the confidence scores and POT threshold values for 1000 scheduling intervals in CAROL. We use the testbed and fault injection module described in Section~\ref{sec:implementation}. The blue bands represent intervals where the confidence fell below threshold value and the model was fine-tuned with the latest collected data.}
    \label{fig:vis}
\end{figure}

\textbf{Confidence Aware Fine-Tuning.} 
After performing optimization in the topology space and obtaining $G_t$ for interval $I_t$, we also evaluate the confidence score $D(M_t, S_t, G_t; \theta)$ (line~\ref{line:confidence} in Alg.~\ref{alg:carol}). We then use the Peak Over Threshold (POT) method~\cite{siffer2017anomaly} that dynamically chooses threshold values below which we fine-tune our model (line~\ref{line:pot}). POT uses extreme value theory to generate a threshold value from a continuous record of past confidence scores. It does this by checking the peak values reached for any period during which values fall below a certain threshold. This threshold is dynamically updated based on incoming data to ensure that the model adapts to non-stationary settings. In any interval, if the confidence score falls below the POT threshold, CAROL fine-tunes the GON network with the latest collected dataset $\Gamma$ (line~\ref{line:tune2}).  A visualization of the training process is shown for a thousand scheduling intervals in Figure~\ref{fig:vis}. The figure clearly demonstrates that the model trains the GON network only when there are dips in the confidence scores, allowing our technique to have much lower fine-tuning overheads compared to other methods that tune their neural networks at each scheduling interval.

\section{Implementation}
\label{sec:implementation}

\begin{figure}[t]
    \centering
    \includegraphics[width=0.65\linewidth]{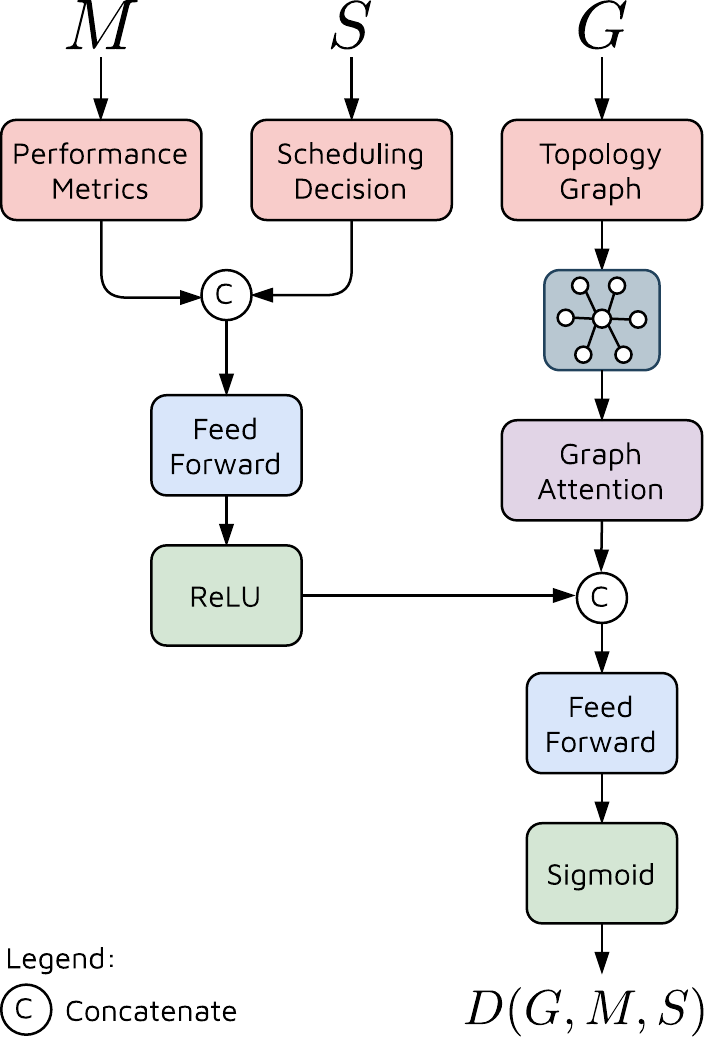}
    \caption{Neural network used in CAROL. The three inputs to the model are shown in red. Feed-forward, graph operations and activations are shown in blue, purple and green.}
    \label{fig:model}
\end{figure}

\subsection{GON Network.} 

It is crucial that the $D$ network is able to capture the correlation between system topology and scheduling decisions with the performance metrics to effectively predict QoS and confidence scores. The model used in our approach is a composite neural network shown in Figure~\ref{fig:model} that infers the correlations between performance metrics, scheduling decision and graph topology to generate the discriminator output. Considering we have $p$ tasks running as a sum of new and active tasks, the scheduling decisions of these are encoded as one-hot vectors of size $|H|$ (number of hosts in the system). We thus get a matrix of scheduling decisions ($S$) of size $[p \times |H|]$. Also, we collect performance metrics of each host $i \in H$ denoted by $M_i$ that includes resource utilization metrics CPU, RAM, Disk, Bandwidth consumption with QoS metrics such as energy consumption and SLO violation rates. We denote the resource utilization metrics of host $i$ as $u_i$ and QoS metrics as $q_i$. We also include the task resource utilization consumption with SLO deadlines, denoted as $t_i$. Thus, $M_i = [u_i, q_i, t_i]$. The $M_i$ vectors for all hosts are stacked together to form the matrix $M$. To encode inputs $M$ and $S$, we use feed-forward layers to bring down the dimension size of the input and $\mathrm{ReLU}$ activation:
\begin{equation}
    E^{\{M, S\}} = \mathrm{ReLU}(\mathrm{FeedForward}([W, S])).
\end{equation}

The topology of the edge federation is represented as a graph with nodes as edge hosts and edges corresponding to edge groups. All edge workers are connected via undirected edges to their respective broker, and all brokers are connected to every other broker. The resource utilization characteristics of each edge host are then used to populate the feature vectors of the nodes in the graph. We denote the feature vector of host $i$ as $e_i$ that are computed from $u_i$. We encode this graph $G$ using a graph attention network~\cite{gat} for scalable computation over the input graph. The motivation behind using a graph attention network is to allow computation over the graph to be agnostic to the number of nodes in the system topology. Graph attention operation performs convolution operation for each node over its neighbors and uses dot product self-attention to aggregate feature vectors.  Thus, we transform the input graph $G$ using \emph{graph-to-graph} updates as: 
\begin{align}
\begin{split}
    e_i &= \sigma \big( \sum_{j \in n(i)} W^q \cdot \mathrm{tanh} ( W\ u_i + b ) \big),\\
\end{split}
\end{align}
where $W^q$ are the attention weights obtained by the weight matrix $W$~\cite{gat} and $n(i)$ are the neighbors of host $i$ in graph $G$.
The stacked representation for all hosts ($e_i$) is represented as $E^G$. Now, we pass this representation through a feed-forward layer with sigmoid activation as
\begin{equation}
    D(M, S, G; \theta) = \mathrm{Sigmoid}(\mathrm{FeedForward}([E^{\{M, S\}}, E^G])).
\end{equation}
Here, the sigmoid function allows us to bring the output in the range $[0, 1]$, the same as that of the true normalized window. 

\subsection{Objective function.} For generating the objective function $O(M_t)$ as part of $\Omega$ in Alg.~\ref{alg:carol}, we use a convex combination of the energy consumption and SLO violation rates of the system as is commonly used in prior work to estimate system performance~\cite{tuli2021cosco, tuli2020dynamic}. For the energy consumption ($q_i^{energy}$) and SLO violation rates ($q_i^{slo}$) of each host $i$, we generate scores for the complete system
\begin{align}
\begin{split}
    q^{energy} &= \sum_{i \in H} q^{energy}_i,\\
    q^{slo} &= \sum_{i \in H} q^{slo}_i.
\end{split}
\end{align}

We then compute the QoS score as
\begin{equation}
\label{eq:qos}
   O(M) = \alpha \cdot q^{energy} + \beta \cdot q^{slo}.
\end{equation}
This is motivated from prior work~\cite{tuli2021cosco} where $\alpha$ and $\beta$ (such that $\alpha+\beta=1$) are hyper-parameters that can be set by users in line with the application requirements. Typically, in real-life settings, these values are around 0.5~\cite{tuli2021cosco, tuli2020dynamic}. For an energy-constrained setting, a higher $\alpha$ value is used, whereas for latency-critical tasks, a higher $\beta$ value is used.


\subsection{Testbed.} Our testbed consists of 16 Raspberry Pi 4B nodes, 8 with 4GB and another 8 with 8GB RAM each. This allows the setup to have heterogeneous nodes with different memory capacities. We consider the same federated topology as a starting point in our experiments. Here, we have 4 LEIs with brokers as 8GB nodes and workers symmetrically distributed from the 4GB and remaining 8GB nodes. The devices run the Ubuntu-18.04 LTS operating system. All WAN and LAN links were 1 Gbps. To emulate each LEI being in geographically distant locations, we use the \texttt{NetLimiter} tool to tweak the communication latency among brokers nodes using the model described in~\cite{gilly2020modelling}. To emulate the gateway devices sending tasks that affect the load distribution across LEIs, we use the gateway mobility model described in~\cite{looga2012mammoth}.

\subsection{Creating the Training Dataset.}  To generate a normal execution trace as the dataset to train the GON model, we ran the widely-used DeFog benchmark applications~\cite{mcchesney2019defog} on our testbed. Specifically, we use the Yolo, PocketSphinx and Aeneas workloads. We build CAROL as a layer on top of the state-of-the-art surrogate modeling-based scheduler (GOBI)~\cite{tuli2021cosco} and execute these tasks as Docker containers. Our runs are divided into five minute long scheduling intervals. We run these traces for 1000 intervals where we periodically change the graph topology every ten intervals. We collect the performance metrics, scheduling decision and graph topology to generate the training dataset $\Lambda = \{M_t, S_t, G_t\}_{t=0}^{1000}$ with 100 different graph topologies. 

\begin{figure}[t]
    \centering
    \includegraphics[width=0.9\linewidth]{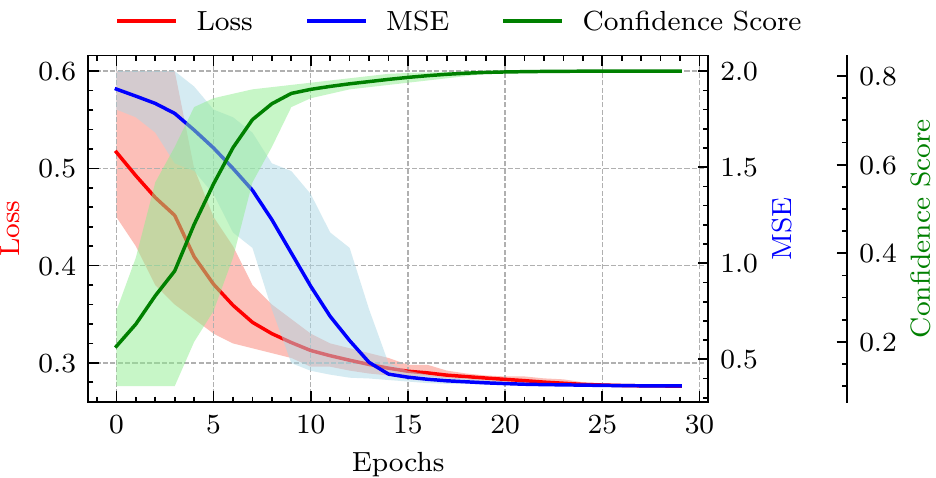}
    \caption{Training plots for the GON model. The model converges in 30 training epochs.}
    \label{fig:convergence}
\end{figure}

\subsection{Training Details and Hyperparameter Selection.} 
For GON, we only change the number of layers in the feed-forward networks in Figure~\ref{fig:model} (keeping a fixed layer size of 128 nodes). We choose the model memory footprint based on grid search so to minimize the Mean-Square-Error (MSE) between the predicted performance metrics and the corresponding values in the dataset. A GON network with a lower parameter size is prone to underfitting and high MSE. A large number of parameters increase the training and fine-tuning time, subsequently increasing the scheduling time due to the slower optimization based generation in~\eqref{eq:opt}. To avoid high MSE values and scheduling times, we use a GON model that consumes $\sim$1GB RAM. A more exhaustive sensitivity analysis is given in Section~\ref{sec:senstivity}. For training, we randomly split the dataset into 80\% training and 20\% testing data. We use a learning rate of $10^{-4}$ with a weight decay of $10^{-5}$ in the \texttt{Adam} optimizer for optimizing the loss function. The training curves showing loss and F1 scores on the test set are shown in Figure~\ref{fig:convergence}. We use the early stopping criterion for convergence. The minibatch size and tabu list size were also obtained using grid search and the values used in our experiments were 32 and 100. More details in Section~\ref{sec:senstivity}.

\subsection{Fault Injection Module.} To generate broker failures at test time, we use an existing fault-injection module~\cite{ye2018fault}.  We create faults of the type: CPU overload, RAM contention, Disk attack and DDOS attack. In a CPU overload attack, a simple CPU hogging application is executed that creates contention of the compute resources. In RAM contention attack, a program is run that performs continuous memory read/write operations. In disk attacks, we run the IOZone benchmark that consumes a large portion of the disk bandwidth. In a DDOS attack, we perform several invalid HTTP server connection requests causing network bandwidth contention. More details are given in~\cite{ye2018fault}. We generate faults using a Poisson distribution with the rate $\lambda_f = 0.5$, sampled uniformly at random from the attack set. These attacks were performed to cause the byzantine failure of broker nodes. 

\subsection{Broker Failure Detection.} In our implementation, all broker nodes periodically (every 30 seconds) \texttt{ping} each other to test which hosts are active in the system. Five  Internet Control Message Protocol (ICMP) packets are sent using the \textit{ping} utility to every other broker and the node waits for a response with a timeout counter set to 10 seconds. If the broker responds, we run the audit checking procedure that verifies the signed log entries since the previous audit~\cite{haeberlen2006case}. If a broker is unresponsive or the audit check fails, an entry is made in a shared PostgreSQL database. For a broker $b \in B$, if all other brokers report it as unresponsive, then $b$ is assumed to be compromised. A `reboot' command is run on this broker in a non-blocking fashion using the \texttt{ssh} utility.

\subsection{Node-shift Implementation.} 
All sharing of resource utilization characteristics across brokers uses the \texttt{rsync} utility. For each worker node, an entry is kept corresponding to the IP address of the broker node to which the worker is assigned. At a node-shift event, the IP address of the new broker is updated for each orphaned node. The worker nodes refresh the broker IP addresses at the start of each scheduling interval. When a worker is assigned as a broker node, it runs the broker management software as a Docker container.  

\subsection{Broker Recovery.} As described in Section~\ref{sec:method}, we assume temporary failures in our setup. Thus, after a broker node fails, it takes 1-5 minutes for this node to restore to the previous state and resume computation/management. To ensure smooth execution of the system, we use the Virtual Router Redundancy Protocol (VRRP) to assign a set of virtual IPs to the broker nodes in the system. We implement this using the \texttt{keepalived} high-availability toolkit.\footnote{Keepalived. \url{https://github.com/acassen/keepalived}. Accessed 7 November 2021.} As soon as a failed node comes back online, we add it to the graph topology and assign it as a worker in the closest active broker as per network latency. This is performed at the time of graph topology initialization at the start of each interval (line~\ref{line:init} of Alg.~\ref{alg:carol}).

\begin{figure*}[!t]
    \centering
    \includegraphics[width=.89\textwidth]{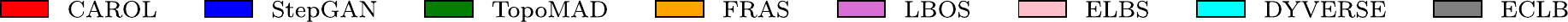}\\
    \includegraphics[width=.79\textwidth]{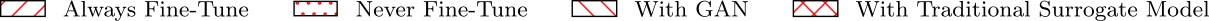}\\
    \subfigure[Energy Consumption]{
    \includegraphics[height=.164\textwidth]{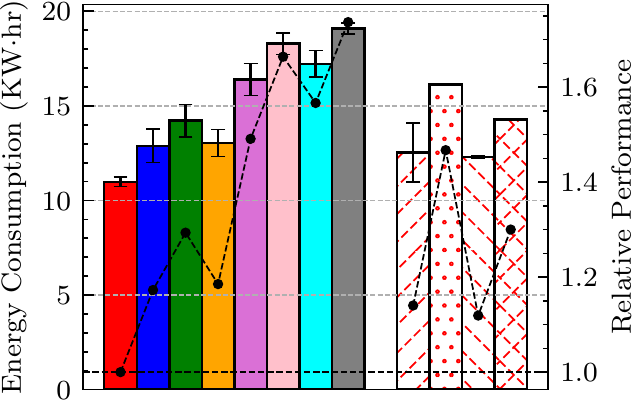}
    \label{fig:energy}
    }
    \subfigure[Response Time]{
    \includegraphics[height=.164\textwidth]{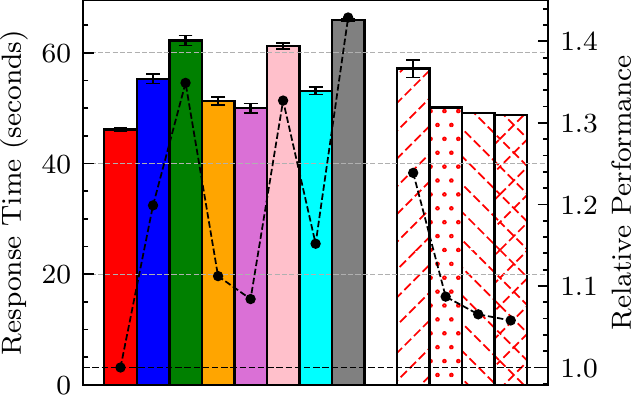}
    \label{fig:rt}
    }
    \subfigure[SLO Violation Rate]{
    \includegraphics[height=.164\textwidth]{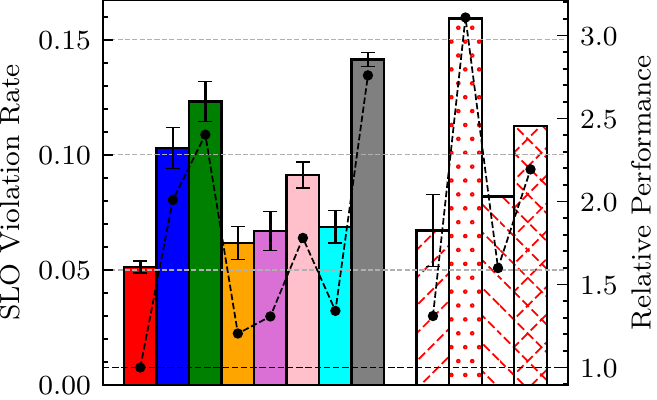}
    \label{fig:slo}
    }\\
    \subfigure[Decision Time]{
    \includegraphics[height=.164\textwidth]{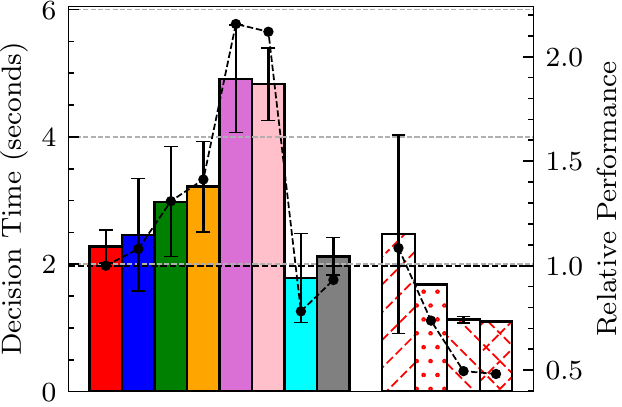}
    \label{fig:decision}
    }
    \subfigure[Memory Consumption]{
    \includegraphics[height=.164\textwidth]{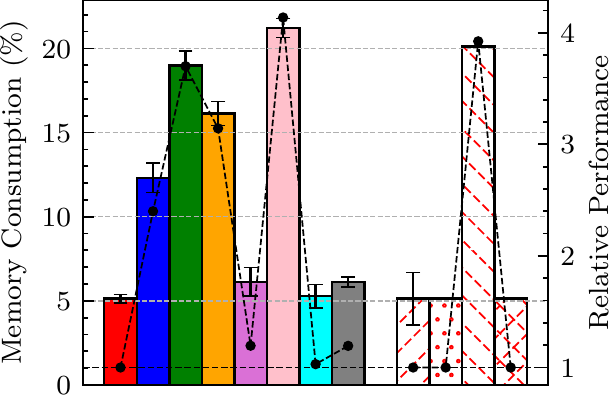}
    \label{fig:mem}
    }
    \subfigure[Fine-Tuning Overhead]{
    \includegraphics[height=.164\textwidth]{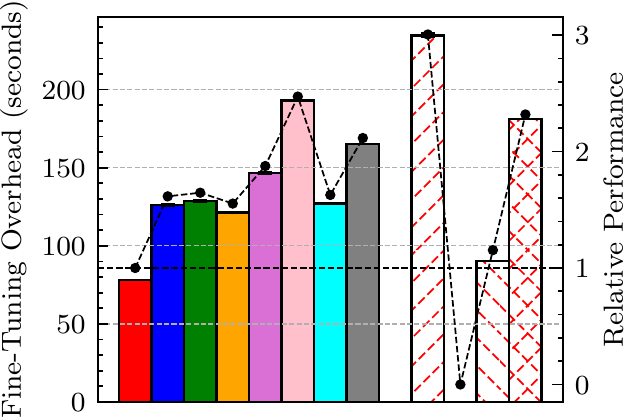}
    \label{fig:overhead}
    }
    \caption{Comparison of CAROL with baseline methods and ablated models. The y-axis on the left shows the absolute values and the one on the right shows relative performance with respect to CAROL. The results corresponding to the ablated models are shown as hatched bar plots.}
    \label{fig:results}
\end{figure*}

\section{Performance Evaluation}
\label{sec:experiments}

We compare the CAROL fault resilience method against a heuristic baseline DYVERSE~\cite{dyverse}, meta-heuristic baseline ECLB~\cite{eclb}, an RL baseline LBOS~\cite{lbos}, two surrogate modelling based methods ELBS~\cite{elbs} and FRAS~\cite{fras}, and two reconstruction models TopoMAD~\cite{topomad} and StepGAN~\cite{stepgan} (more details in Section~\ref{sec:related_work}). As TopoMAD and StepGAN are only fault-detection methods, we supplement them with the priority based load-balancing policy from the next best baseline, \textit{i.e.}, FRAS. We use hyperparameters of the baseline models as presented in their respective papers. We train all deep learning models using the PyTorch-1.8.0~\cite{paszke2019pytorch} library. 

\subsection{Workloads.} To test the generalization capability of the various resilience models and their adaptive capacity, we do not use the DeFog benchmarks as workloads at test time. Instead, we use the \textit{AIoTBench} applications~\cite{luo2018aiot, tuli2022splitplace}. AIoTBench is an AI-based edge computing benchmark suite that consists of various real-world computer vision application instances. The seven specific application types correspond to the neural networks they utilize. These include three typical heavy-weight networks: ResNet18, ResNet34, ResNext32x4d, as well as four light-weight networks: SqueezeNet, GoogleNet, MobileNetV2, MnasNet.\footnote{AIoTBench: \url{https://www.benchcouncil.org/aibench/aiotbench/index.html}. Accessed: 5 November 2021.} This benchmark is used due to its volatile utilization characteristics and heterogeneous resource requirements. The benchmark consists of 50,000 images from the COCO dataset as workloads~\cite{coco}. To evaluate the proposed method in a controlled environment, we abstract out the users and IoT layers described in Section~\ref{sec:method} and use a discrete probability distribution to generate tasks as container instances. Thus, at the start of each scheduling interval, we create new tasks from a Poisson distribution with rate $\lambda_t = 1.2$, sampled uniformly from the three applications. The Poisson distribution is a natural choice for a bag-of-tasks workload model, common in edge environments~\cite{mao2016dynamic}. Our tasks are executed using Docker containers. We run all experiments for 100 scheduling intervals, with each interval being five minutes long, giving a total experiment time of 8 hours 20 minutes. We average over five runs and use diverse workload types to ensure statistical significance of our experiments.

\subsection{Evaluation Metrics.} We measure the energy consumption of the federated setup, average response time and SLO violation rate of completed tasks. We consider the relative definition of SLO (as in~\cite{tuli2021cosco}) where the deadline is the 90$^{th}$ percentile response time for the same application on the state-of-the-art method StepGAN that has the highest F1 scores among the baselines. We also consider the average inference time of the models, that is the time to decide the steps for system resilience, such as deciding the required node-shift operations. We also compare the memory consumption of the techniques and the fine-tuning overhead averaged over the scheduling intervals.

\subsection{Comparison with Baselines.}
We now present results comparing CAROL with baselines. For our experiments, we use $\alpha = \beta = 0.5$ in~\eqref{eq:qos} as per prior work~\cite{tuli2020dynamic, tuli2021cosco}. Figure~\ref{fig:energy} compares the total energy consumption of all models and shows that CAROL is able to execute tasks with minimum energy consumption in the federated edge environment. The presence of the average energy consumption metric of all the edge nodes within the QoS score calculation as shown in~\eqref{eq:qos} forces the model to take energy-efficient node-shift decisions. It results in the minimum number of active hosts with the remaining hosts in standby mode to conserve energy. CAROL reduces energy consumption by 16.45\% compared to StepGAN, the model having minimum energy consumption across all baseline methods. 

Figure~\ref{fig:rt} shows the average response time per task in seconds. The response time is measured as the time difference between the timestamps of task creation and result generation and is a crucial metric to compare the service throughput offered by the edge federation. The figure demonstrates that CAROL reduces this metric by 8.04\% compared to the best baseline FRAS. 

Figure~\ref{fig:slo} shows the fraction of SLA violations out of the total completed tasks in the 100 scheduling intervals. The figure shows that CAROL has the lowest SLO violation rate of only  5.12\%, 17.01\% lower than the least violation rates among the baselines, \textit{i.e.}, 6.17\% of the FRAS method. Allowing node-shift at each interval with the SLO violation rate in the QoS score calculation ensures that the topology minimizes response time and subsequently the SLO violation rates on an average. Moreover, CAROL is given the SLO deadline for each task that it utilizes to decide the number of worker nodes to allocate per task to minimize the SLA violation metric in the QoS score. 

Figure~\ref{fig:decision} shows the comparison of average time to decide the fault-tolerance steps for each model. The figure shows that CAROL has a lower decision time compared to all AI-based methods. LBOS and ELBS have the highest decision time due to the time-consuming weighted round-robin and match-making algorithms used in these methods. The least decision time is of the DYVERSE heuristic method. However, CAROL has only 6.77\% higher decision time than DYVERSE.

Figure~\ref{fig:mem} shows the average percentage memory utilization of the fault-tolerance models. The surrogate and generative methods have high memory consumption. ELBS has the highest memory consumption due to the resource intensive fuzzy neural networks in this approach. The RL based method, LBOS, has the lowest memory footprint among the AI baselines due to a lightweight Q-table used in the Q learning model in LBOS. Compared to the heuristic methods, CAROL has memory consumption ($\sim$5\%), thanks to the memory-efficient GON model.  

Figure~\ref{fig:overhead} shows the overheads for fine-tuning the AI models and dynamically updating the priority scores in the heuristic models. The overhead is measured for the complete experiment, \textit{i.e.}, over 100 scheduling intervals. The figure shows that CAROL has the lowest overhead of 78.12 seconds, only 0.78 seconds on an average per five-minute interval. This is 35.62\% lower than the baseline with the lowest overhead, \textit{i.e.}, FRAS that takes 121.35 seconds to update its model periodically. The significantly lower overhead of CAROL is due to the economic confidence-based model fine-tuning. 

\begin{figure*}[!t]
    \centering
    \subfigure[Learning Rate]{
    \includegraphics[height=.171\textwidth]{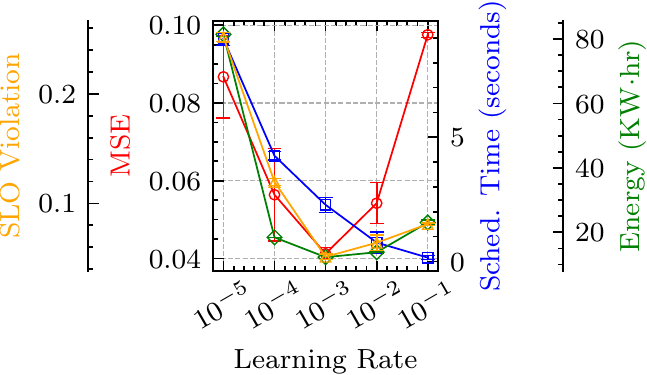}
    \label{fig:lr}
    }
    \subfigure[Memory Size]{
    \includegraphics[height=.171\textwidth]{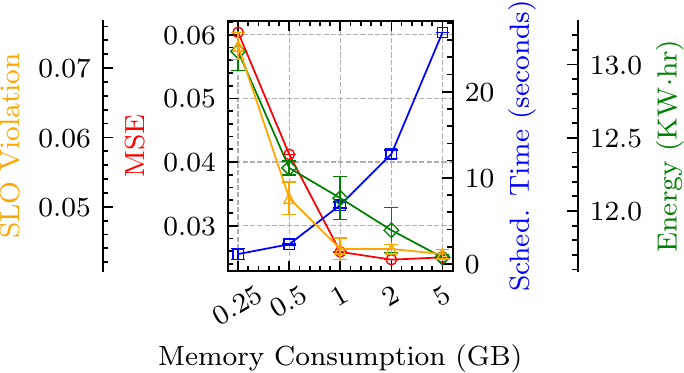}
    \label{fig:mems}
    }
    \subfigure[Tabu List Size]{
    \includegraphics[height=.171\textwidth]{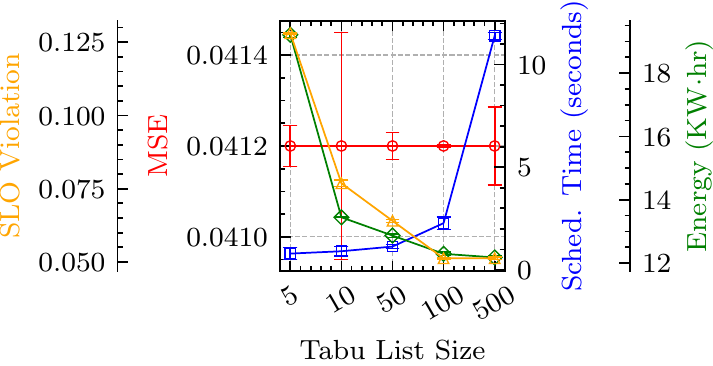}
    \label{fig:tabu}
    }
    \caption{Sensitivity Analysis.}
    \label{fig:sens}
\end{figure*}

Overall, we observe that the ability to approximate the QoS scores for a given graph of the system topology and scheduling decision enables the GON-based generative model to predict the confidence scores and accurately predict QoS scores for any given state of the system. This, with the node-shifting techniques in CAROL, allows us to optimize the QoS of the federated environment.

\subsection{Ablation Analysis}

To study the relative importance of each component of the model, we exclude every major component one at a time and observe how it affects the performance of the scheduler. An overview of this ablation analysis is given in Fig~\ref{fig:results} as hatched bars. First, we consider the CAROL model without the confidence scores, fine-tuning the GON network at every scheduling interval (\textit{Always Fine-Tune} model). We also consider CAROL without fine-tuning the GON network in any interval (\textit{Never Fine-Tune} model). Next, we consider a model with a traditional GAN network instead of the GON based confidence and QoS prediction (\textit{With GAN} model). Lastly, we consider the CAROL model with a traditional feed-forward surrogate network instead of GON (\textit{With Traditional Surrogate} model). We report the following findings:

\begin{itemize}[leftmargin=*]
    \item Without the confidence-aware model fine-tuning, the overheads significantly increase, causing a higher average decision time and response time of applications. 
    \item Without ever fine-tuning the model, it does not adapt in non-stationary settings giving rise to poor QoS scores.
    \item Without the GON model, and using a GAN instead, the decision time is reduced as we do not need to optimize in the topology space. However, the memory consumption increases from 5\% to 30\%, which impacts the running applications and broker management. This causes the SLO violation rate to increase by 6\%.
    \item Without the GON model, and using a traditional feed-forward model, the decision time is reduced again, but at the cost of higher fine-tuning overheads.
\end{itemize}
We observe that the confidence-aware model fine-tuning and using the GON network have the maximum impact on performance values, accounting to nearly 70\% of the total impact.

\subsection{Sensitivity Analysis}
\label{sec:senstivity}

Figure~\ref{fig:sens} provides a sensitivity analysis of the performance of the CAROL model with learning rate, \textit{i.e.}, $\gamma$ in \eqref{eq:opt}, memory consumption and size of the tabu list. This sensitivity analysis highlights the trade-off between the QoS scores and scheduling time as we increase either of these parameters. The variation of performance metrics with the learning rate is more straightforward. The scheduling time of CAROL decreases as we increase the learning rate. This is due to larger jumps in the optimization steps. However, for a high learning rate of $\gamma \geq 10^{-2}$, the model is unable to converge to the optima and hence we see the increase in the MSE scores, energy consumption and SLO violation rates for these values. The QoS scores are best for the learning rate of $\gamma = 10^{-3}$ and hence have been used in our experiments.

As the number of layers in the GON model increases, so does the memory footprint. This increases the scheduling time as it takes longer to generate samples by running optimization in the input space. However, a higher layer count improves the prediction performance by giving lower MSE scores, leading to lower energy consumption and SLO violation rates due to the more accurate QoS prediction. However, we see marginal improvement in energy consumption and SLO violation rates for memory consumption $>$1GB, but large increase in the scheduling time of the model. Thus, we use the model with 1GB of memory footprint (3 layers) in our experiments. Finally, increasing the size of the tabu list increases the scheduling time and gives better energy and SLO scores. We use a 100 size tabu list in our experiments.

\section{Conclusions and Future Work}
We have developed a novel method for resilient computing in edge federations. Our method uses a lightweight generative network as a surrogate model to efficiently and precisely map performance metrics like energy consumption and SLO violation rates for a given graph topology and scheduling decision. In the event of a broker failure, CAROL reactively optimizes the graph topology to accommodate the orphaned worker nodes. The topology is chosen by running a tabu search to optimize the QoS scores predicted by the surrogate model. CAROL uses the discriminator output as a confidence score that allows us to run model fine-tuning steps only when confidence scores drop below running thresholds. This parsimonious fine-tuning gives rise to 35.6\% lower overheads compared to the current state-of-the-art. Performance evaluation of a real edge test-bed with AI and IoT based benchmark applications show that the low overheads in CAROL can improve energy consumption and SLO violation rates by 16.5\% and 17.0\% compared to the state-of-the-art.

The current approach is able to achieve optimal performance due to its lower fine-tuning overheads compared to prior work. Our methods are best suited to highly dynamic systems. For stationary settings, we propose to extend the current reactive model to a proactive scheme that is able to prevent node failures. However, proactive optimization may entail higher computation for improved predictive performance. Such challenges would be addressed in the future.

\section*{Software Availability}
The code and relevant result reproduction scripts are available at \url{https://github.com/imperial-qore/CAROL}.



\bibliographystyle{IEEEtran}
\bibliography{references}

\end{document}